%% file: merged-submit_v2.tex
\documentclass[aps,prl,twocolumn,showpacs,superscriptaddress]{revtex4} \usepackage{graphicx,amssymb,amsfonts,amsmath}

\usepackage{ulem}
\usepackage{subfiles}
\usepackage{hyperref}

\newcommand{\br}{\mathbf{r}}
\newcommand{\bk}{\mathbf{k}}
\newcommand{\etal}{ \text{et al.}, } 
\newcommand{\CO}{(color online)\;}

\newcommand{\jn}{\mathbf{j}_n}
\newcommand{\je}{\mathbf{j}_e}
\newcommand{\bj}{{\bf{j}}}
\newcommand{\bv}{{\bf{v}}}

\newcommand{\intdk}{\frac{1}{4\pi^2}\int\!\!d^2\bk^{\phantom{2}}\,}

\begin{document}

\subfile{gravity_v11_tomerge.st}

\clearpage

\subfile{gravity_supplementary_tomerge.st}

\end{document}

%% file: gravity_v11_tomerge.st
\title{ Interacting fermionic atoms in optical lattices diffuse symmetrically upwards and downwards in a gravitational potential} \author{Stephan Mandt} \affiliation{Institute for Theoretical Physics, University of Cologne, 50937 Cologne, Germany}
\author{Akos Rapp} \affiliation{Institute for Theoretical Physics, University of Cologne, 50937 Cologne, Germany} \author{Achim Rosch} \affiliation{Institute for Theoretical Physics, University of Cologne, 50937 Cologne, Germany} \date{\today}


\begin{abstract} We consider a cloud of fermionic atoms in an optical lattice described by a Hubbard model with an additional linear potential. 
While  homogeneous interacting systems mainly show damped Bloch oscillations and heating, a finite cloud behaves differently:
It  expands symmetrically  such that  gains of
potential energy at the top are compensated by losses at the bottom. Interactions stabilize the necessary heat currents by inducing gradients of the inverse temperature $1/T$, with $T<0$  at the bottom  of the cloud. An
analytic solution of hydrodynamic equations shows that the width of the cloud  increases with $t^{1/3}$ for long times 
consistent with results from our Boltzmann simulations. 
\end{abstract}
	


\maketitle
Measuring the conductivity is probably the most fundamental experiment when investigating the properties of a metal.
The influence of constant forces arising from electric or gravitational fields on quantum particles in periodic potentials is an old and well studied problem \cite{bloch0}. For free particles, Bragg scattering from the periodic potential induces Bloch oscillations \cite{bloch0,bloch1,expBloch} . While difficult to observe in solids due to disorder and interactions, such Bloch oscillations have been measured in semiconductor superlattices \cite{expSC} and for ultracold bosonic atoms in optical lattices \cite{bloch1,expBloch} created by standing waves of lasers, e.g., to determine the masses of atoms with high precision \cite{masses}.

Recent experimental developments make it also possible to load fermionic ultracold atoms in the lowest band of  optical lattices. 
Using an equal population of two hyperfine levels and Feshbach resonances (e.g., of $^{40}K$) it became possible to realize  the Hubbard model (see below)  with tunable interaction strength. While the temperatures in current experiments are still rather high, it was possible to see \cite{mottEsslinger,mottBloch} signatures of the onset of a metal-insulator transition induced by strong interactions. Constant external forces can be realized using either gravitation
or accelerated lattices \cite{bloch1} (after a careful elimination of other confining potentials as in Ref.~\cite{expansion}).

The Hubbard model in an electric or gravitational field has attracted previously considerable attention \cite{aoki,Mott-tDMRG,freericks1,freericks2,blochBose,eckstein,prelovsek,bosonexpansiontwo}, partially motivated by the question how large electric fields can lead to a breakdown of the Mott insulating state. 
When discussing the physics of such systems either for weak or strong interactions, it
is important to take energy conservation into account. While real solids are usually probed in contact with some thermal bath, ultracold atoms provide almost ideal realizations of closed quantum systems  implying severe restrictions on the dynamics. For a translationally invariant, infinite system one expects that even weak interactions lead to a damping of the Bloch oscillations (at least in the absence of superfluidity and in dimensions larger than 1). During this process, potential energy is converted into heat. As long as there is no coupling to an external bath which can transport the heat away, the system gets hotter and hotter. It finally reaches a steady state characterized by an infinite temperature and vanishing current. For moderately strong interactions we expect that the same steady state is reached, but Bloch oscillations may become overdamped. Finally, in the Mott regime, the currents and dissipation are exponentially suppressed for small fields \cite{aoki,eckstein}, but in the long-time limit heating up to $T=\infty$ is expected. While we are not aware of a study which carefully tests this plausible picture, it seems
to be consistent with the available numerical results, e.g. the short-time behavior extracted from dynamical mean-field equations studied by Eckstein, Oka and Werner~\cite{eckstein}.

\begin{figure}
\begin{center}
\includegraphics[width=0.55 \linewidth]{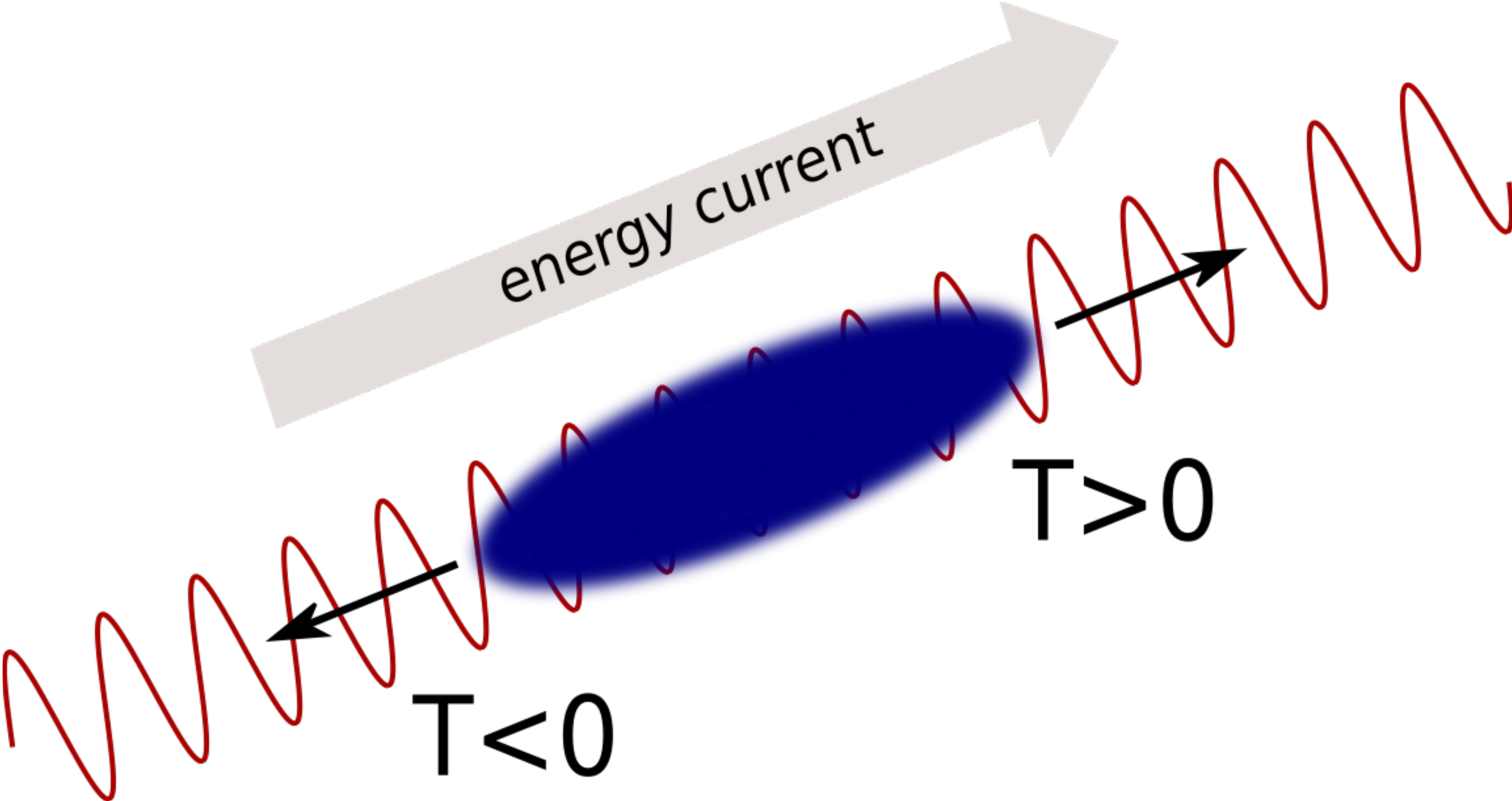}
\end{center}
\caption{\CO Atomic cloud in an optical lattice in the presence of gravity. A symmetric expansion of the cloud is possible by transporting energy ``uphill'' as atoms lose potential energy at the bottom and gain it at the top. This energy current is associated with  a gradient of $1/T$ with $T<0$ ($T>0$) at the bottom (top) of the cloud. \label{fig1}}
\end{figure}

Here we argue that the physics of an inhomogeneous system, i.e., a finite atomic cloud, 
is rather different; see Fig. \ref{fig1}. 
For bosons, this question has been studied using the Gross-Pitaevskii equation \cite{bosonexpansiontwo,blochBose}. The motion of the system is governed by energy conservation and
the fact that the kinetic and interaction energies per atom are bounded not only from below but also from above (assuming that the optical lattice is sufficiently deep that interband transitions can be safely neglected). 
 Energy conservation prohibits that the cloud as a whole
moves up or down over long distances. For noninteracting particles this implies that only Bloch oscillations  are possible. With interactions a second possibility
arises: energy conservation allows that the cloud expands \textit{symmetrically}, losing potential energy at the bottom and gaining it at the top of the cloud. To this end, energy has to be transported over macroscopic distances upwards. We can expect that the interacting system explores this part of the phase space: the cloud expands. For a sufficiently large cloud and not too strong driving
forces, one can expect that most of the system is approximately in local equilibrium such that a local temperature can be defined. The presence of an energy current and the absence of a particle current in the center of the cloud implies a gradient in
temperature or rather in $\beta=1/T$. Combining this insight with the expectation that $T \to \infty$ in the center of the cloud for long times (as in the homogeneous system), we obtain the situation sketched in Fig.~\ref{fig1}:
The cloud expands, using an energy current driven by a gradient of $\beta=1/T$ with $T>0$ at the top of the cloud and \textit{negative} absolute temperatures, $T<0$, at the bottom. The numerical confirmation of this qualitative picture by a Boltzmann simulation and the associated quantitative theory presented below are the main results of this Letter. Note that negative $T$, i.e. systems where high-energy states have a higher occupation, $e^{- E/k_B T}$, than 
low-energy states,
are well-defined for closed systems with an upper bound in energy.
In Ref.~[\onlinecite{negativeT}] we discuss how $T<0$ can be realized and detected.

To substantiate our picture, we study the dynamics of a cloud  governed by a two-dimensional (2D) fermionic Hubbard model in the presence of a gravitational force, 
\begin{eqnarray} 
H&=&-J \sum_{\langle ij \rangle, \sigma} c^\dagger_{i\sigma} c_{j\sigma}+ U \sum_i n_{i\uparrow} n_{i \downarrow}+\sum_{i,\sigma} V_i n_{i\sigma}, \label{eq:H0} \\ 
V_i &=& g r_i^x,   \nonumber
\end{eqnarray} 
with hopping rate $J$ and interaction $U$. The number of atoms of spin $\sigma$ at site $i$ located at position ${\bf r}_i$ is given by $n_{i\sigma}= c^\dagger_{i\sigma} c_{i\sigma}$. We set the lattice constant $a = 1$ in the following. The constant force $g$ is applied in the $x$ direction, and this term prohibits equilibrium at finite $T$. As we are mainly interested in the dynamics in this direction, we consider for simplicity  a translationally invariant system in the $y$ direction.  Initially the system is in equilibrium at a given $T=J$ (a typical temperature for current experiments), confined by an extra harmonic trap in the $x$-direction, which is switched off at $t=0$.  For $g=0$, the Hamiltonian (\ref{eq:H0}) has been realized 
in Ref.~[\onlinecite{expansion}] where the expansion dynamics was studied also theoretically.
 Considering a 2D instead of a 3D setup has the advantage that in experiments it allows one to change the value of $g$ by tilting the 2D lattice. While $g \approx J$ can be reached easily in experiments~\cite{expansion}, we are more interested in a regime where $g$ is much weaker. As an alternative to tilting, one can also use accelerated lattices \cite{bloch1}.

\begin{figure}
\begin{center}
\includegraphics[angle=270,width= \linewidth,clip=true]{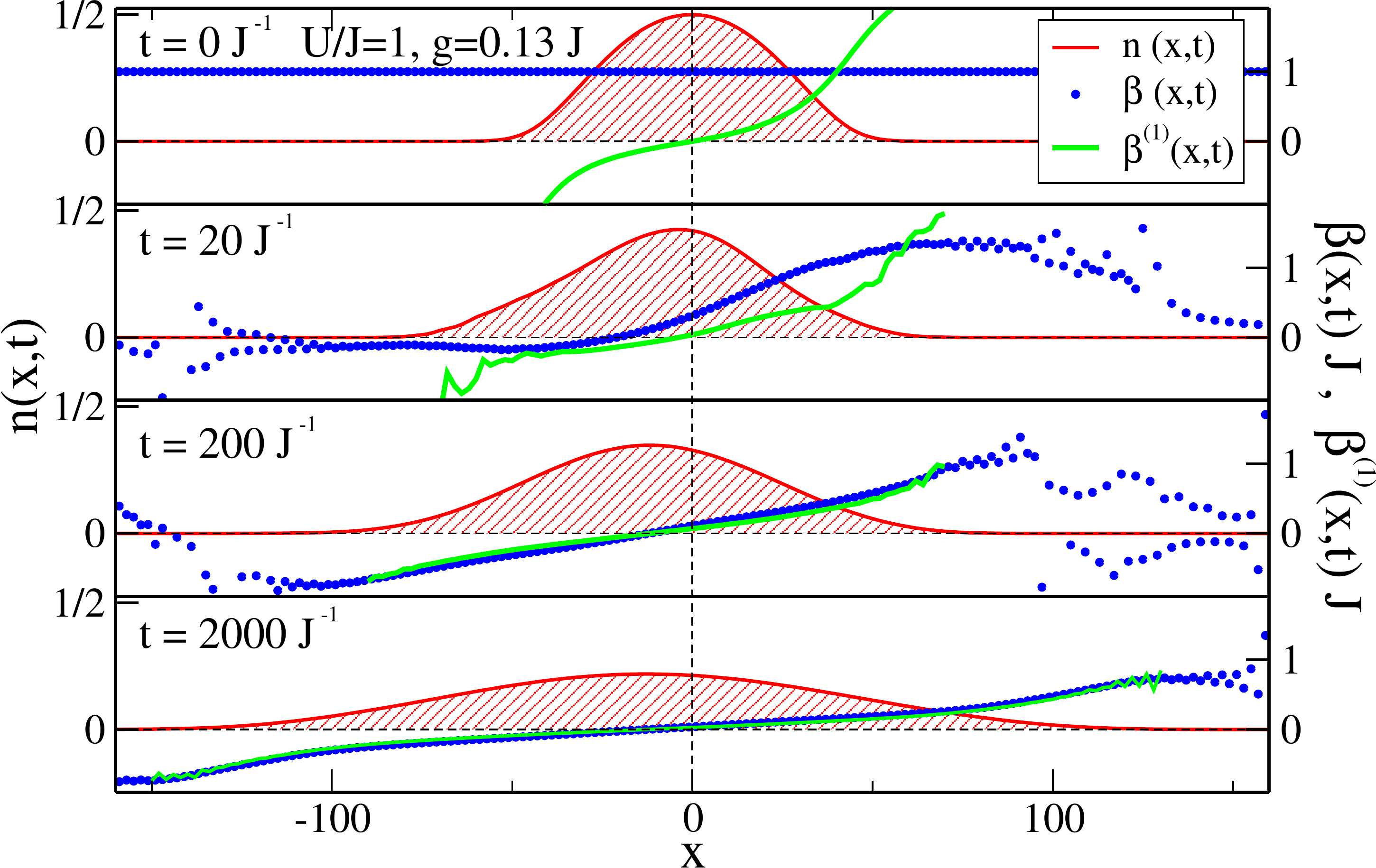}
\end{center}


\caption{\CO Densities $n(x,t)$ (red) and inverse local temperatures $\beta(x,t) = 1/T(x,t)$ (blue dots) from the Boltzmann simulation for $U=J$ and $g\approx 0.13 J$. The  expansion is associated with
 a $1/T$ gradient with $T<0$, $T=\infty$, $T>0$ at the bottom, in the center and at the top of the cloud, respectively. For long $t$, $\beta$ approaches
 $\beta^{(1)}=- \partial_x n/(n \partial_x V)$ (green).
In the tails, local temperatures cease to be a useful concept and the noise reflects Bloch oscillations. 
 \label{fig2}}
\end{figure}

For not too large values of $U$ and $g$, the system can be described by a semiclassical Boltzmann
equation. As for an inhomogeneous system this is a numerically demanding integro-differential equation, we employ a version of the relaxation 
time approximation \cite{expansion} where both particle number and energy is conserved:
\begin{equation}
    \partial_t f + \mathbf{v}_\bk \cdot \nabla_\br f + \mathbf{F} \cdot \nabla_\bk f = -\frac{1}{\tau(n,e)}( f - f^0_{n,e}) \;,
\label{eq:Boltzmann}
\end{equation}
Here $f(\br,\bk,t)$ is the occupation probability in phase space and
the force term $\mathbf{F} = - \nabla_\br (g\, x) - U \nabla_\br
n(\br,t)$ contains the external potential and interaction corrections on
the Hartree level. The velocity is defined as $\mathbf{v}_\bk = \nabla_\bk\epsilon_{\bf k}$ with the dispersion relation $\epsilon_{\bf k} = -2 J [\cos( k_x) + \cos(k_y)] $.  The parameters $T(\br,t)$ and $\mu(\br,t)$ of the Fermi function $f^0_{n,e}$ are determined from $n=\frac{1}{4 \pi^2} \int d^2{ k} \, f^0(\epsilon_{\bf k}-\mu,T) $ and $e=\frac{1}{4 \pi^2} \int d^2{ k} \, \epsilon_{\bf k}  f^0(\epsilon_{\bf k}-\mu,T)$, where $n=n_{\uparrow}=n_{\downarrow}=n(\br,t)$ and $e=e(\br,t)$ are the local particle and kinetic energy densities per spin, respectively. We use
a scattering rate $1/\tau(n,e) \sim U^2$ which arises from interparticle scattering with momentum transfer to the lattice (umklapp) and which was determined in Ref.~\cite{expansion} to reproduce the $e$- and $n$-dependent diffusion
constant of the Hubbard model to order $U^2$ (obtained from an independent calculation). In the regime of high $T$, relevant for our study, $1/\tau(n,e)\approx  n (1-n)/\tau_0$  is approximately independent of temperature with  $1/\tau_0 \approx 0.609\, U^2/J$ in $D=2$ (see Ref.~\cite{expansion}
for more details).

\begin{figure}
\begin{center}
\includegraphics[width=1. \linewidth,clip=true]{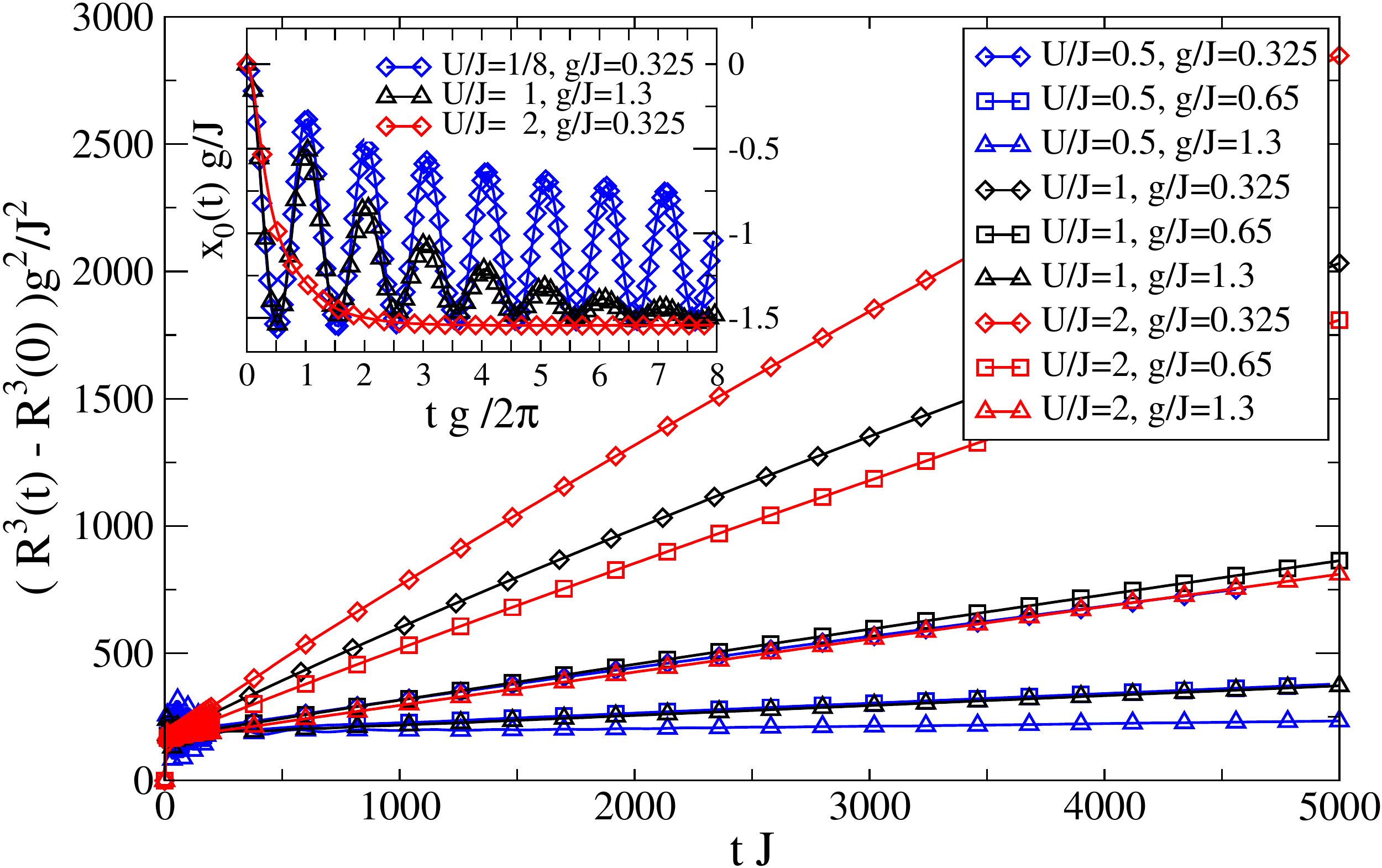}
\end{center}


\caption{\CO  To a good approximation, the cloud radius cubed grows linearly in time, $R^3(t)\sim (t-t_0)$, for a wide range of $U$ and $g$.
 Inset: Center of mass of the cloud $x_0(t)$ showing damped Bloch oscillations for weak $U$.
 \label{fig3}}
\end{figure}

A typical result of the Boltzmann equation is shown in Fig.~\ref{fig2}. In the gravitational potential, the cloud falls initially until energy- and temperature profiles become antisymmetric relative to the center of the cloud
such that $T = \infty$ in the center.  For the parameters of Fig.~\ref{fig2}, Bloch oscillations of the center $x_0(t) \sim \int\!\!dx\; x n$ are overdamped but become visible for weaker interactions; see  inset of Fig.~\ref{fig3}. Note that there are always Bloch oscillations in the tails of the cloud where low densities prohibit scattering. The most prominent feature is, however, the symmetric expansion of the cloud in the long-time limit. The width of the cloud increases slowly; see  Fig.~\ref{fig3}. For the local temperatures (defined above) one observes the formation of a gradient in $1/T$ with negative (positive) absolute temperatures at the bottom (top) of the cloud.  As discussed above, this  gradient is associated to the heat currents needed for the symmetric expansion of the cloud.

To obtain an analytic theory of the effects described above, we derive the hydrodynamic equations for $n({\bf r},t)$, and $e({\bf  r},t)$ in the regime where the scattering time is smaller than the Bloch oscillation period, $g  \tau/2\pi \ll 1$. The opposite limit of weakly damped Bloch oscillations is briefly discussed in the conclusions. We use the standard approach to expand the Boltzmann Eq.~(\ref{eq:Boltzmann}) to linear order 
around the local equilibrium distribution $f^0_{e,n}(\epsilon_{\textbf{k}})$. As we are interested in the dynamics for long times, we use that $T$ will become very high and $n$ small with
\begin{equation}
 T = -4 J^2 n /e  \label{eq:beta0}
\end{equation}
to leading order. Using also that $1/\tau\approx n/\tau_0$  in this limit, we obtain (see supplement\cite{supplement}) 
\begin{eqnarray}
\partial_t n&+&\nabla \jn=0, \qquad \partial_t e+\nabla \je=-\jn \nabla V \label{Diffeq} \\
\jn&=&- \frac{J^2 \tau_0}{n} \left(2 + \frac{e^2}{16 J^2 n^2} \right) \nabla n+\frac{ \tau_0 e}{2 n}  \nabla V + \frac{ \tau_0 e}{8 n^2} \nabla e  \nonumber\\
\je&=&- \frac{3 J^2 \tau_0}{2 n} \nabla e+\frac{3 \tau_0 e^2}{8 n^2}  \nabla V  \label{jn}
\end{eqnarray} 
where $\jn$ and $\je$ are particle and energy currents. Above we have included all subleading
terms in the high-$T$ expansion, which contribute in the $t \to \infty$ limit according to the analysis below. Kinetic energy is created by Joule heating with the rate $ - \jn \nabla V(x)$ where $V(x)=g x + U n(x)$ includes Hartree corrections. 

Note that the hydrodynamic equations can be also derived directly from the Hubbard model. Up to changes in numerical prefactors, Eqs.~(\ref{Diffeq}) and (\ref{jn}) therefore should be \textit{exact} for the large $T$ limit of the  Hubbard model in dimensions $D>1$ for large clouds and weak potentials $V$ (the breakdown of the hydrodynamic equations found for $g=0$ in \cite{expansion} is not important here). In that sense, our analysis and results below do \textit{not} rely on our specific Boltzmann equation. Also note that the 1D case is more subtle due to the integrability of the Hubbard model.

\begin{figure}
\begin{center}
\includegraphics[angle=270,width=\linewidth,clip]{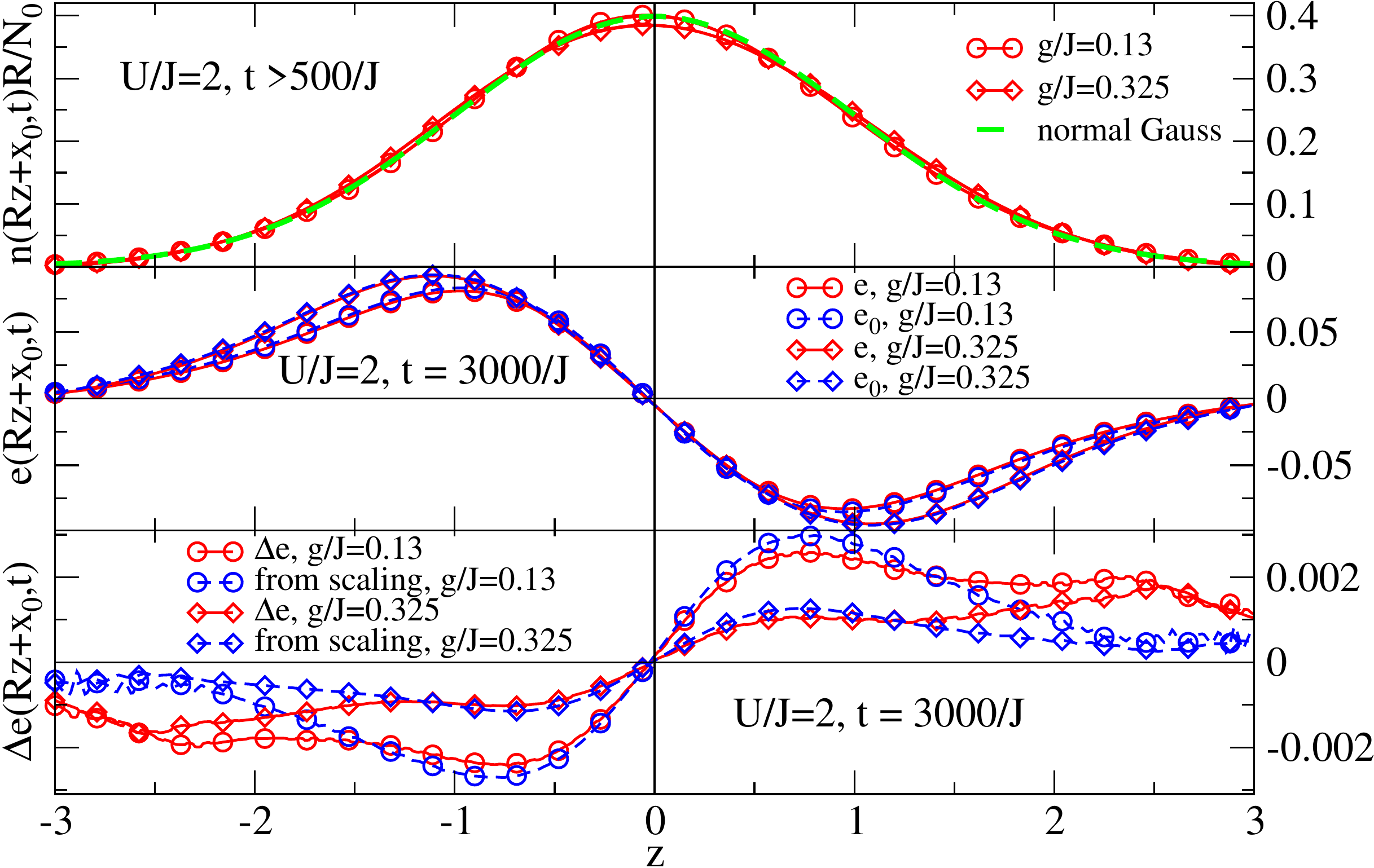}
\end{center}


\caption{\CO Top: Normalized density as a function of $z=x/R(t)$ for  $t J =500,1500, 2500, 3500$. As predicted by Eq. (\ref{eq:solveF}) the shape is approximately Gaussian.  Middle: Energy density $e$ compared to $e_0$ given by Eq.~(\ref{eansatz}). Bottom: Comparison of the subleading correction, $\Delta e$, to  Eq.~(\ref{eq:Deltae}) where $j_n^x$ and $n$ are taken from our simulation.
 \label{fig4}}
\end{figure}

Surprisingly, it is possible to analyze the long-time limit of the nonlinear, coupled partial differential Eqs.~(\ref{Diffeq}) and ~(\ref{jn})
analytically. Measuring distance $x$ relative to the center off mass, we start from the scaling ansatz
\begin{eqnarray}
n(x,t)=N_0 \frac{1}{R(t)} F[x/R(t)]\;, \label{eq:n-ansatz}
\end{eqnarray}
where $F$ is a dimensionless scaling function of unit width and $\int\!\!dz\;F[z]=1$ such that $N_0=\int\!\!dx\; n$, $R^2(t)=\int\!\!dx\;x^2 n/N_0$. Our goal is to calculate both $R(t)$ and $F[z]$ in the long-time limit where $R(t)$ is large. On the top panel of Fig.~\ref{fig4} we show the scaling function $F$ obtained by rescaling the density profiles from the Boltzmann simulation according to Eq.~(\ref{eq:n-ansatz}). To obtain a scaling ansatz for $e$, the most important step is to realize that \textit{to leading order}, energy conservation prohibits expansion. We therefore set $e=e_0+\Delta e$, and determine $e_0$ by setting $j_n^x=0$ in Eq.~(\ref{jn}) for $e=e_0$ (neglecting subleading terms $\propto e^2$). This leads to the ansatz
\begin{eqnarray}\label{eansatz}
e_0=\frac{4 J^2}{\partial_x V} \partial_x n, \qquad e=e_0+\frac{J}{R(t)^{1+\gamma}} G[x/R(t)]
\end{eqnarray}
with an unknown exponent $\gamma>1$ (such that $\Delta e \ll e_0$) and a new scaling function $G$. To check whether $e \approx e_0$ to leading order approximation, we  compare  $1/T$ from our numerical data to Eq.~(\ref{eq:beta0}) with $e\approx e_0$
and find excellent agreement (see Fig.~\ref{fig2}). While this provides already a precise, quantitative theory for the temperature gradient, only \textit{corrections} to $e_0$ determine how the cloud expands.

From the particle number continuity equation in Eq.~(\ref{Diffeq})  we can calculate
$j_n^x=-\int^x \partial_t n(x',t) dx'$ $=N_0 x F  R^{-2}\dot R$ using Eq.~(\ref{eq:n-ansatz}), which has to be equivalent to $j_n^x$ in Eq.~(\ref{jn}). We can match the scaling dimension of this term to the subleading terms proportional to $\Delta e$  if  $R^{\gamma-1} \dot R=const.$, implying that
\begin{eqnarray}
R(t)&=& \alpha_0\,    \left( \frac{ \tau_0 J^4 }{N_0 g^2}\right)^{1/\gamma}(t-t_0)^{1/\gamma}
\end{eqnarray}
where $\alpha_0$ and $t_0$ are integration constants and the other prefactors of $R(t)$ have been chosen for later convenience. 

We can now substitute these results into the energy continuity equation, Eq.~(\ref{Diffeq}), from which we obtain terms proportional to $t^{-1},t^{-3/\gamma},t^{-1-2/\gamma}$ and $t^{-2-1/\gamma}$. From the condition that the two leading terms have to cancel to fulfill energy conservation, we obtain
\begin{eqnarray}
\gamma&=&3,\; R \sim t^{1/3}  \label{scaling}  \\
\partial_z^3 && \!\!\!\!\!\!\!\!\!\!\!  \log  F[z]= \frac{\alpha_0^3}{18} z F[z],\; F[z]\approx \frac{e^{-z^2/2}}{\sqrt{2 \pi}}+O(\alpha_0^3) \label{eq:solveF}\\
G[z]&=&\frac{2 N_0 J^3}{ g ^{3}}\left( \frac{\alpha_0^3  z F[z]^2}{3} + 
\frac{ F'[z]^3}{ F[z]^2}-\frac{2 F'[z] F''[z]}{ F[z]} \right)\nonumber \\ \label{gsol}  
\end{eqnarray}
Therefore, we conclude that the width of the cloud increases with $t^{1/3}$ for long times  in the diffusive regime. In comparison, subdiffusive exponents $\gamma^{-1} \le 1/4$ \cite{bosonexpansiontwo} and $\gamma^{-1} = 0.19 $ \cite{blochBose} have also been obtained for bosons within time-dependent mean-field theory.  The parameter $\alpha_0$ and therefore the shape of $F$ depend on the initial conditions but for our simulations we find that $\alpha_0^3/18$ is  small and therefore the rescaled density profile $F$ is close to a Gaussian according to  Eq.~(\ref{eq:solveF}). 

Figure~\ref{fig4} shows that the analytic approach describes the results of the Boltzmann simulations with high precision.
We combine Eqs.~(\ref{eansatz}) and (\ref{gsol}) to predict 
\begin{eqnarray}
\Delta e\approx \frac{2}{g \tau_0} n j_n^x + \frac{2 J^4 (\partial_x n)^3}{g^3 n^2} - \frac{4 J^4  n (\partial_x n) (\partial_x^2 n)}{g^3 n^2} \label{eq:Deltae}
\end{eqnarray}
Testing this relation numerically, we find good agreement even for this subleading term; see Fig. \ref{fig4}.
As expected, our hydrodynamic equations are, however, not valid in the tails of the cloud where $1/\tau$ is small.

In conclusion, we have shown that interactions  lead to a symmetric expansion of fermionic 
clouds in optical lattices 
subject to a constant force. In the limit where the scattering time is short compared to a Bloch oscillation period, $\tau \ll t_B=2 \pi/g$, we derived analytically a precise quantitative theory for the expansion of the cloud based on hydrodynamics. It is an interesting question for further studies to investigate analytically also the opposite limit, $\tau \gg t_B$, where diffusion constants are proportional to $1/\tau$ instead of $\tau$ as only  scattering allows the atoms to move forward. Nevertheless, energy conservation will again lead to a symmetric expansion, however, with reduced speed. For $t \to \infty$ one reaches this regime for arbitrary $U$  as for $t \gg N_0^4/(\alpha_0^3 g J^4 \tau_0^4)$ one always obtains $\tau \gg t_B$. For moderately strong interactions, however, we expect that atomic losses will make it very difficult to track experimentally the system for such long times.

We acknowledge financial support by SFB TR 12 and SFB 608 of the DFG and the German National Academic Foundation and useful discussions with 
I. Bloch, E. Demler, V. Gurarie, M. Kunze, S. Manmana, A. M. Rey, U. Schneider, and R. Sch\"utzhold.

%% file: gravity_supplementary_tomerge.st
\title{Supplementary material to ``Interacting fermionic atoms in optical lattices diffuse symmetrically upwards and downwards in a gravitational potential''} 
\author{Stephan Mandt} \affiliation{Institute for Theoretical Physics, University of Cologne, 50937 Cologne, Germany}
\author{Akos Rapp} \affiliation{Institute for Theoretical Physics, University of Cologne, 50937 Cologne, Germany} \author{Achim Rosch} \affiliation{Institute for Theoretical Physics, University of Cologne, 50937 Cologne, Germany} \date{\today}

\begin{abstract}

\end{abstract} 

\maketitle

\section{The motion of the center of mass of the cloud}

The center of mass of the cloud is defined by
\begin{equation}
	x_0(t) = \frac{1}{N_0} \int\!\!dx\; x n(x,t)\;,
\end{equation}
where $N_0 = \int\!\!dx\; n(x,t)$. Using energy conservation,
\begin{eqnarray}
	\int\!\!dx \left( e(x,t) + g x n(x,t) + \frac{1}{2} U n^2(x,t) \right) = \rm{const.}
\end{eqnarray}
we can estimate $x_0(t)$ for very short and very long times. With $x_0(t=0)=0$, the center of mass of the cloud at time $t$ is simply given by
\begin{eqnarray}
	x_0(t) &=& \frac{1}{ N_0 g} \int\!\!dx ( e(x,0) -  e(x,t)  ) \nonumber \\
          && + \frac{U}{2 N_0 g}\int\!\!dx ( n^2(x,0) -  n^2(x,t) )  \;.\label{eq:x0t}
\end{eqnarray}

As is shown in Fig.~\ref{energies} for short times, the total Hartree energy decays only very slowly (see below). For short (or moderately long) times, we can therefore approximate it as a constant. For the case of weakly damped Bloch oscillations 
it is useful to examine Eq.~(\ref{eq:x0t}) for $t = \pi/ g$, i.e., after half of a Bloch period. In the non-interacting case, for all particles $k_x(t=\pi/g) = k_x(t=0) + \pi $ while  $k_y(t=\pi/g) = k_y(t=0)$. In $d=2$ this implies $e(x,t=\pi/g) = 0 $ as the two terms in the kinetic energy cancel each other. As a consequence,
\begin{equation}
 x_0(t=\pi/g) \approx \tilde x_0 \equiv \frac{1}{ N_0 g} \int\!\!dx \; e(x,0)
\end{equation}
also for weak interactions. For high initial temperatures $T(t=0) \gg J$ we can approximate $e \approx -4 J^2 n/T$ and obtain $\tilde x_0 \approx  -4 J^2 /( T g)$.
For $T(t=0) = J$ used in our simulations, one gets $\tilde x_0\approx - 1.59 J/g$, which describes the numerical results to good approximation, see the inset in Fig. 3. of the main text.

The same arguments can be applied to $t_n = (2 n +1)\pi/g, n \in Z$ for weakly damped Bloch oscillations. This explains
the form of $x_0(t)$ shown in the inset of Fig. 3 (main text) for small $U$ and/or large $g$, see also Fig.~\ref{energies}: While the amplitude of the Bloch oscillations is damped, the local minima of $x_0(t)$, are at an approximately constant value, $x_0(t_n)\approx \tilde x_0$.
Therefore, on the time scales where on the one hand the Bloch oscillations have died off and on the other hand the change of the Hartree energy can still be neglected, we obtain 
\begin{equation}
	x_0(t) \approx \tilde x_0.
\end{equation}

Formally, the Hartree energy vanishes for $t \to \infty$ and therefore
\begin{equation}
	x_0(t\to \infty) = \tilde x_0+ \frac{U}{2 N_0 g}\int\!\!dx  \, n^2(x,0).
\end{equation}
This asymptotic value is, however, reached only very slowly as $\int\!\!dx\, n^2\propto 1/R(t) \propto 1/t^{1/3}$ in the regime where our hydrodynamic approach
is valid.

For the validity of the scaling ansatz used in the main text, where an antisymmetric form of $e(x,t)=-e(-x,t)$ was used, it is important to check that
the total kinetic energy vanishes much faster than the Hartree energy which is indeed the case, see Fig.~\ref{energies}.

\begin{figure}[ht]
	\includegraphics[width=.48\textwidth,clip]{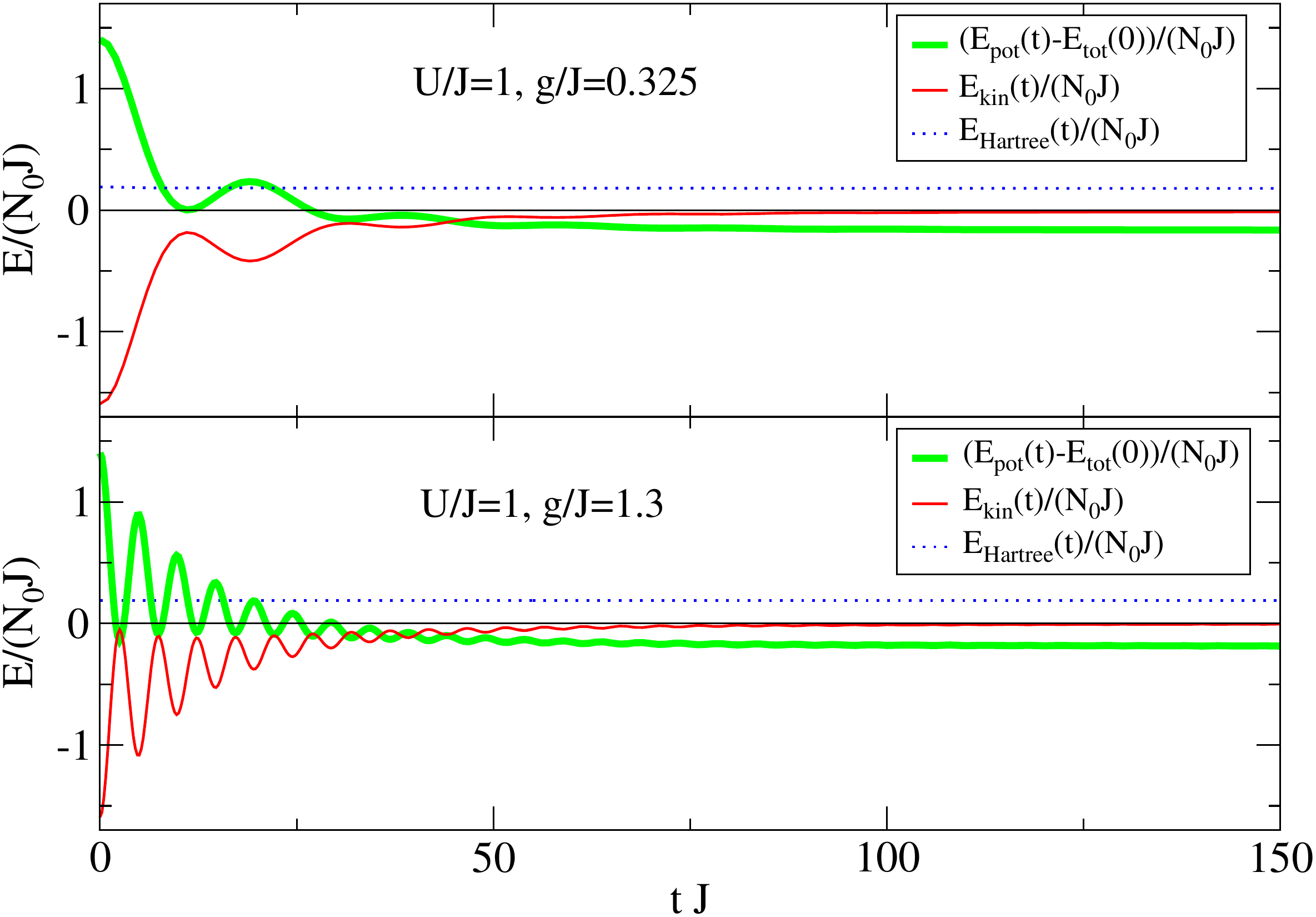}
 	\caption{\CO Total kinetic energy, potential energy, and Hartree energy (per particle) for different times as a function of $t$ for $U/J = 1 $, $g/J = 0.325 $ and $g/J = 1.3 $. 
	 }
	\label{energies}
\end{figure}

\section{Derivation of hydrodynamic equations from the Boltzmann equation}

In this section we shall derive hydrodynamic equations from the Boltzmann equation
\begin{equation}
	\partial_t f + \bv_\bk \cdot \nabla_\br f - (\nabla_\br V ) \cdot \nabla_\bk f = -\tau^{-1}(n,e) ( f - f^0) \label{def:eq:Boltzmann-supp}
\end{equation}
for weak potentials $V$ and large scattering rates $\tau^{-1}$ for high temperatures. The local densities and currents per spin component are defined by
\begin{eqnarray}
	n &=& \intdk f \,,\; e = \intdk \epsilon_\bk f \,,\label{eq:def:n} \\
	\bj_n &=&  \intdk \bv_\bk f \,,\; \bj_e =  \intdk \bv_\bk \epsilon_\bk f \,. \label{eq:def:j}
\end{eqnarray}

We start from the particle number and kinetic energy continuity equations,
\begin{eqnarray}
	\dot n &=& - \nabla_\br \bj_n \;, \nonumber \\
	\dot e &=& - \nabla_\br \bj_e + (-\nabla_\br V) \cdot \bj_n \;.\label{def:eq:diffusion}
\end{eqnarray}
Our goal is to express the currents $\bj_n$ and $\bj_e$ as functions of $n$ and $e$ in lowest orders in $e$.

In the diffusive limit, where $\tau g \ll 1$, we can solve the Boltzmann equation iteratively to get
\begin{eqnarray}
	f &\approx& f^0 + (-\tau)( \bv_\bk \cdot \nabla_\br f^0  + (-\nabla_\br V) \cdot \nabla_\bk f^0 ) \;, \label{eq:Boltzmann_iteration}
\end{eqnarray}
where 
\begin{equation}
	f^0 = \frac{1}{\zeta(\br,t) e^{\beta(\br,t) \epsilon_\bk } + 1 }
\end{equation}
with the fugacity $\zeta$ and inverse temperature $\beta$ chosen such that
\begin{eqnarray}
	n &=& \intdk f^0 \;,\; e = \intdk \epsilon_\bk f^0 \;.
\end{eqnarray}
Note that although one should also consider the term  $ (-\tau) \partial_t f^0[n(\br,t),e(\br,t),\epsilon_\bk]$ in Eq.~(\ref{eq:Boltzmann_iteration}), it does not contribute to the currents due to the odd power of $\bv_\bk$ in the corresponding momentum integral in Eq.~(\ref{eq:def:j}).

For high temperatures $\beta J \ll 1$, the local kinetic energy density $e$ is small compared to the bandwidth. Thus we expand the distribution function in $\beta J$,
\begin{eqnarray}
	f^0 &=& \frac{1}{1+\zeta }-\frac{\zeta }{(1+\zeta )^2}  \epsilon_\bk \beta 
	+\frac{(-1+\zeta ) \zeta }{2 (1+\zeta )^3}  \epsilon_\bk^2 \beta ^2 \nonumber \\
	&& -\frac{\zeta  (1+(-4+\zeta ) \zeta )}{6 (1+\zeta )^4} \epsilon_\bk^3\beta ^3 +{\cal O}(\beta ^4) \;,
\end{eqnarray}
which we substitute in the definitions of $n$ and $e$. We can easily perform the momentum integrals, and find
\begin{eqnarray}
	n &=& \frac{1}{1+\zeta }+ A_2(d) \frac{(-1+\zeta ) \zeta  }{2 (1+\zeta )^3} \beta^2 \,, \\
	e &=& - A_2(d)\frac{\zeta }{(1+\zeta )^2} \beta - A_4(d) \frac{\zeta  (1+(-4+\zeta ) \zeta )}{6 (1+\zeta )^4}  \beta^3  \nonumber \,,
\end{eqnarray}
where $A_n(d) = \intdk \epsilon_\bk^n$, e.g., $A_2(d=2) = 4 J^2$ and $A_4(d=2) = 36 J^4$. We can solve these equations for $\zeta$ and $\beta$ as a function of $n$ up to order $e^3$. Substituting these results back to $f^0$ gives

\vspace{1mm}

\begin{widetext}
\begin{eqnarray}
	f^0 &\approx& n + \frac{ e \epsilon_\bk  }{A_2(d)} - \frac{e^2 (1-2n)(A_2(d)- \epsilon_\bk^2 )}{2 A_2^2(d)(1-n)n} +\frac{ e^3 (1- 6 (1-n) n) \epsilon_\bk \left(\epsilon_\bk^2 - A_4(d)/A_2(d)\right)}{6 A_2(d)^3 (1-n)^2 n^2} \;.
\end{eqnarray}
We use this expression in Eq.~(\ref{eq:Boltzmann_iteration}) to get $f = f[n,e]$. Then from Eq.~(\ref{eq:def:j}) we obtain the currents by keeping terms up to ${\cal O}(e^2)$:
\begin{eqnarray}
	\bj_n &=&  -\tau \; C \; \nabla_\br n + \tau \; \frac{C}{A_2(d)}  \; (\nabla_\br V)  \; e \nonumber \\
	&& -\tau   \frac{A_2(d) C - B_2(d) }{2 A_2^2(d)} \frac{  (1 - 2 n + 2 n^2 )}{ (1-n)^2 n^2} ( \nabla_\br n) e^2 
           +\tau   \frac{A_2(d) C - B_2(d) }{ A_2^2(d)} \frac{ (1-2 n) }{ (1-n) n}  e \nabla_\br e \;; \\
	\bj_e &=& -\tau \; \frac{B_2(d)}{A_2(d)} \; \nabla_\br e  + \tau \; \frac{B_2(d)}{A_2^2(d)} \frac{1-2n}{(1-n) n}  (\nabla_\br V)  e^2 \;.
\end{eqnarray}

\end{widetext}
Here we used that $\intdk v_\bk^i v_\bk^j = \delta_{ij} C; C  =   2 J^2 a^2$ and $\intdk \epsilon_\bk^2 v_\bk^i v_\bk^j = \delta_{ij} B_2(d); B_2(d=2) = 6 J^4 a^2$. Note that the first term in Eq.~(\ref{eq:Boltzmann_iteration}) does not contribute to the currents since $f^0$ corresponds to an equilibrium distribution function. Thus the currents are always proportional to $\tau \approx \tau_0 /n$. 

For the hydrodynamic equations in the main text, we only keep terms which contribute for $t \to \infty$.